\documentclass[aps,twocolumn,groupeaddress]{revtex4}

\newcommand{\me}{\mathrm{e}}

\newcommand{\dif}{\mathrm{d}}

\usepackage{graphicx}

\usepackage{amsmath}

\usepackage{amsfonts}

\begin{document}
\title{Role of fluctuations in a snug-fit mechanism
of KcsA channel selectivity}
\author{D. Asthagiri}
\author{Lawrence R. Pratt}
\affiliation{Theoretical Division, Los Alamos National Laboratory, Los
Alamos, NM 87545}

\author{Michael E. Paulaitis}
\affiliation{Department of Chemical Engineering, Ohio State
University, Columbus, OH  43210, USA}
\date{\today}
\begin{abstract}
\end{abstract}
\maketitle

\section{Introduction}

{\bf The KcsA potassium channel belongs to a class of K$^+$ channels
that is selective for K$^+$ over Na$^+$ at rates of K$^+$ transport
approaching the diffusion limit \cite{cmiller:jmembiol83}. This
selectivity is explained thermodynamically in terms of favorable
partitioning of K$^+$ relative to Na$^+$ in a narrow selectivity filter
in the channel. One mechanism for selectivity based on the atomic
structure of the KcsA channel  \cite{rmk:science98} invokes the size
difference between K$^+$ and Na$^+$, and the molecular complementarity
of the selectivity filter with the larger K$^+$ ion \cite{armstrong:72}.
An alternative view holds that size-based selectivity is precluded
because atomic structural fluctuations are greater than the size
difference between these two ions \cite{roux:nature04}. We examine these
hypotheses by calculating the distribution of binding energies for
Na$^+$ and K$^+$ in a simplified model of the selectivity filter of the
KcsA channel. We find that Na$^+$ binds strongly to the selectivity
filter with a mean binding energy substantially lower than that for
K$^+$. The difference is comparable to the difference in hydration free
energies of Na$^+$ and K$^+$ in bulk aqueous solution. Thus, the average
filter binding energies do not discriminate Na$^+$ from K$^+$ when
measured from the baseline of the difference in bulk hydration free
energies. Instead, Na$^+$/K$^+$ discrimination can be attributed to
scarcity of good binding configurations for Na$^+$ compared to K$^+$. 
That relative scarcity is quantified as enhanced binding energy
fluctuations, and is consistent with predicted relative constriction of
the filter by Na$^+$.}

\begin{figure}[b]
\begin{center}
\includegraphics[width=3in]{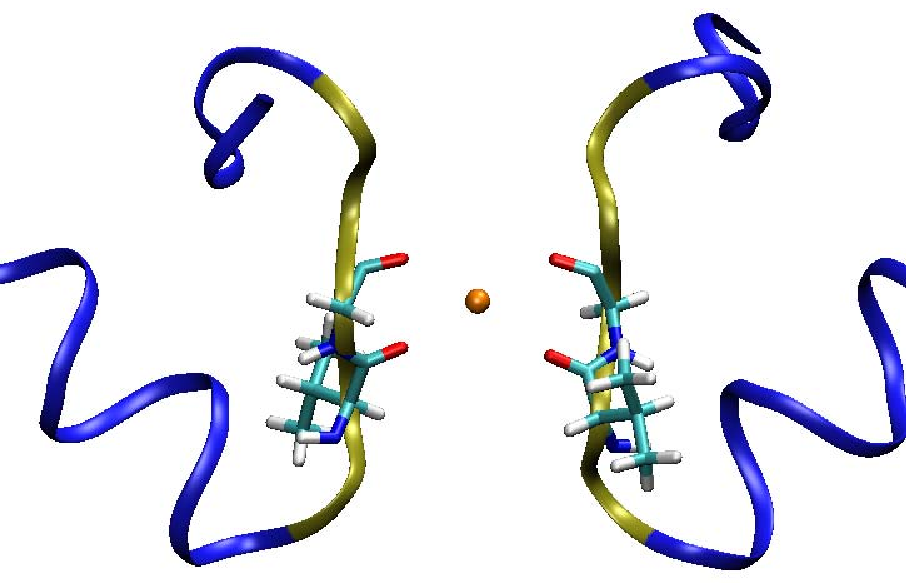}
\end{center}
\caption{Depiction of the channel filter. Only two of the four chains are
shown for clarity. Heavy atoms in the blue ribbon segment experience a
harmonic mean field, \emph{i.e.,} they are ``restrained''  with spring
of force constant $k$ = 10~kcal/mol-\AA$^2$. The segment colored yellow
is the TTVGYG sequence of amino acids comprising the selectivity filter. The S$_2$ site
 defined by VG is shown with the bound ion.}
\label{fg:model}
\end{figure}

A tetrameric constellation of four TTVGYG amino acid sequences comprises
the selectivity filter shown in Fig.~\ref{fg:model}.  This K$^+$
signature sequence is strongly conserved across a variety of K$^+$
channels \cite{rmk:science98}. In each binding site, for example the
site defined by (VG)$_4$, the ion is coordinated by four carbonyl
oxygens from above and four from below. This binding site provides a
snug fit for K$^+$. Thus it is argued that when K$^+$ enters the binding
site its dehydration is compensated by favorable interactions with the
carbonyl oxygens \cite{rmk:science98,rmk:febs03,armstrong:re03}. In
contrast, the smaller Na$^+$ ion is unable to interact optimally with
all of the available carbonyls. This imperfect compensation of
dehydration is suggested as the basis for the observed selectivity.
Implicit in this hypothesis is the idea that  the channel is stiff,
and poor coordination of all the carbonyl oxygens leads to {\em
weaker\/} binding of the ion to the binding site. We will refer to this
as the snug-fit mechanism.

A focused attempt to evaluate the snug-fit mechanism at a molecular
level is the recent work of Noskov~\emph{et~al.} \cite{roux:nature04}.
They argued that size-based selectivity is precluded because atomic
structural fluctuations are greater than the difference in Pauling radii
for the two ions. Through numerical experimentation involving extensive
free energy calculations, they concluded that local interactions leading
to structural flexibility of the binding site provided a key to
selectivity for K$^+$. They pointed-out that the selectivity of the
binding site is sensitive to carbonyl-carbonyl repulsive interactions. 
Artificially turning-off carbonyl-carbonyl electrostatic repulsion 
while retaining carbonyl-ion electrostatic attractions was found to
shift the thermodynamic selection in favor of Na$^+$.

In the analysis below we follow Noskov~{\em et~al.} \cite{roux:nature04}
in studying a simplified model of the selectivity filter. A distinction
of the present work with that of Noskov~{\em et~al.} is the use of more
parsimonious statistical thermodynamic analyses. This sharpens the
physical points that may be observational. On this basis we might expect
experimental validation of the key result of these analyses,
\emph{i.e.,} that constriction of the filter and enhanced fluctuations
accompany each other in the binding of an under-sized ion such as
Na$^+$. Further experimental investigation of the joint association of
these factors might help to clarify the combined roles of fluctuations
and a snug fit \cite{rmk:science98,roux:nature04} in the mechanism of
ion selectivity by the KcsA channel.

\section{Theory}

The equilibrium selectivity of the filter can be characterized  by the
difference in the interaction free energy for transferring a Na$^+$ ion from
water  into the selectivity filter compared to the case for a K$^+$
ion. Thus we study
\begin{multline}
\Delta\mu^\mathrm{ex} =
\left\lbrack \mu_\mathrm{Na^+}^\mathrm{ex}\left(\mathrm{filter}\right) - 
\mu_\mathrm{K^+}^\mathrm{ex}\left(\mathrm{filter}\right)\right\rbrack  \\  
 -   \left\lbrack \mu_\mathrm{Na^+}^\mathrm{ex}\left(\mathrm{aq}\right)- \mu_\mathrm{K^+}^\mathrm{ex}\left(\mathrm{aq}\right)\right\rbrack
 \\
\equiv \Delta\mu^\mathrm{ex}\left(\mathrm{filter}\right) - 
\Delta\mu^\mathrm{ex} \left(\mathrm{aq}\right)~.
\label{eq:selec}
\end{multline}
Here $\mu_\mathrm{X}^\mathrm{ex}(\mathrm{aq})$ (X=K$^+$, Na$^+$) is the
hydration free energy of the ion, and $\mu_\mathrm{X}^\mathrm{ex}(\mathrm{filter})$ is the analogous quantity
in the selectivity filter.    $\mu_\mathrm{X}^\mathrm{ex}(\mathrm{aq})$, the excess chemical  potential, is
that  part of the  chemical potential that would vanish if intermolecular interactions were to be
neglected.  Thus, $\mu_\mathrm{X}^\mathrm{ex}(\mathrm{aq})$ is understood to be
referenced to the ideal gas result at the same density and temperature. 
This technical point deserves emphasis because we could adopt a standard
state in which $\Delta\mu^\mathrm{ex} \left(\mathrm{aq}\right)$ would
vanish, but that would not change any physical consideration.

The potential distribution theorem \cite{lrp:book,lrp:apc02}  
\begin{equation}
\me^{\beta \mu_\mathrm{X}^\mathrm{ex}} = \int \me^{\beta \varepsilon} P_\mathrm{X}(\varepsilon) \dif\varepsilon = \langle  \me^{\beta \varepsilon} \rangle
\label{eq:pdt1}
\end{equation}
tells us how $\mu_\mathrm{X}^\mathrm{ex}$ may be calculated. Here
$\varepsilon$ is the binding energy of the X ion to the medium;
$P_\mathrm{X}\left(\varepsilon\right)$ is the probability density
function  of this interaction energy.
$P_\mathrm{X}\left(\varepsilon\right)$ is generated with the ion and the
medium fully coupled at temperature $T= 1/k_\mathrm{B}\beta$ where $
k_\mathrm{B}$ is Boltzmann's constant.  In the present approach ion positions contributing to the sample
are those corresponding to its natural motion in the ion-protein system. 
If $P_\mathrm{X}(\varepsilon)$ is well-described by a gaussian of mean
$\left\langle\varepsilon\right\rangle$ and variance $\sigma^2$ =
$\left\langle\left(\varepsilon
-\left\langle\varepsilon\right\rangle\right)^2\right\rangle$, then
$P_\mathrm{X}(\varepsilon) \propto \me^{-(\varepsilon
-\left\langle\varepsilon\right\rangle)^2/ 2\sigma^2}$,  and
\begin{equation}
\mu_\mathrm{X}^\mathrm{ex} = \left\langle\varepsilon\right\rangle + \beta \sigma^2/ 2 ~.
\label{eq:work1}
\end{equation}
The width  parameter $\sigma$ gauges a fluctuation contribution.  

Note that in sampling the fully-coupled system the term $\beta \sigma^2/ 2$
raises the chemical potential above the mean binding energy
$\left\langle \varepsilon \right\rangle$. This should be contrasted with 
the case of sampling from the uncoupled subsystems.  In the uncoupled
case, contributions beyond a mean field term \emph{lower} the free energy.
The distinction reflects the fact that the mean binding energies are computed from different probability distributions. $P_\mathrm{X}^{(0)}\left(\varepsilon\right)$ is the probability distribution function for the binding energy when the ion and the medium are  uncoupled, and is given by
\begin{eqnarray}
P_\mathrm{X}^{(0)}\left(\varepsilon\right) = 
\me^{\beta\left(\varepsilon - \mu_\mathrm{X}^\mathrm{ex}\right)}
P_\mathrm{X}\left(\varepsilon\right)~.
\label{eq:pointwise}
\end{eqnarray}

Intrinsic fluctuations of binding energies are associated with these probability distributions, and contribute to
the free energies in a natural way, 
\begin{eqnarray}
\left\langle \varepsilon \right\rangle & = & \int\varepsilon{P_\mathrm{X}\left(\varepsilon\right)}d\varepsilon 
\nonumber \\
& = &  \mu_\mathrm{X}^\mathrm{ex} -k_{\mathrm{B}}T\int{P_\mathrm{X}\left(\varepsilon\right)}\ln{\frac{P_\mathrm{X}\left(\varepsilon\right)}{P_\mathrm{X}^{(0)}\left(\varepsilon\right)}}d\varepsilon~.
\label{eq:aveps}
\end{eqnarray}
which uses Eq.~\ref{eq:pointwise}. Here the additional fluctuation contribution, the right-most term in Eq.~\ref{eq:aveps}, suggests an entropic contribution to the free energy of binding beyond
the mean interaction energy for the joint system. Note specifically, however, that $\left\langle \varepsilon \right\rangle$ is expected to be temperature dependent so that this additional fluctuation contribution
is not an identification of the thermodynamic entropy contribution.

Additional physical perspective can be obtained from the 
formal relation 
\begin{eqnarray}
\mu^\mathrm{ex}_\mathrm{X} =
kT \ln\int\limits_{-\infty}^{\bar{\varepsilon}}P_\mathrm{X}\left(
\varepsilon\right)\me^{\beta\varepsilon }\dif\varepsilon 
-kT\ln\int\limits_{-\infty}^{\bar{\varepsilon}}P_\mathrm{X}^{(0)}\left(
\varepsilon\right)\dif\varepsilon~,
\label{eq:sep0}
\end{eqnarray}
which is true independently of the binding energy cutoff parameter
$\bar{\varepsilon}$; see 
for example \cite{lrp:book,CPMS}.
We expect $P_\mathrm{X}\left(\varepsilon\right)$ to be 
concentrated near $\varepsilon\approx\left\langle\varepsilon\right\rangle$,
and thus it is natural to choose $\bar\varepsilon$ so that 
\begin{eqnarray}
\left\langle\varepsilon\right\rangle = kT \ln\int\limits_{-\infty}^{\bar{\varepsilon}}P_\mathrm{X}\left(
\varepsilon\right)\me^{\beta\varepsilon }\dif\varepsilon~.
\end{eqnarray}
 Then
\begin{eqnarray}
- \left(\frac{\mu^\mathrm{ex}_\mathrm{X} -
\left\langle\varepsilon\right\rangle}{kT}\right) = 
\ln\int\limits_{-\infty}^{\bar{\varepsilon}}P_\mathrm{X}^{(0)}\left(
\varepsilon\right)\dif\varepsilon~.
\label{eq:sep1}
\end{eqnarray}
The natural estimate is $\bar\varepsilon\approx \mu^\mathrm{ex}_\mathrm{X} $.

\begin{figure}
\begin{center}
\includegraphics[width=3in]{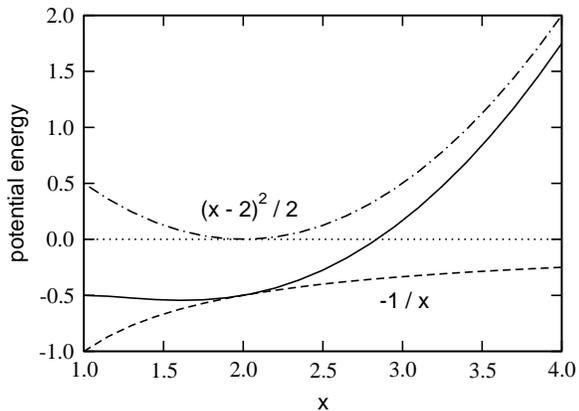}
\end{center}
\caption{ An example how coupling an ion and the
filter might enhance fluctuations.   Here a coordinate in a parabolic
potential-energy ($-\cdot-$) well also experiences a coulombic attraction 
($- -$) displacing the minimum energy position leftward (unbroken line). The fluctuations
observed for the fully-coupled system,  corresponding to the potential
energy function given by the solid curve, characterizes the available
states on the  harmonic potential function in the neighborhood of the
physical binding energies. The curvature at the minimum is reduced,
fluctuations are enhanced, and the observed fluctuations raise the free
energy above the mean binding energy, according to Eq.~\ref{eq:work1}.}
\label{fg:pushpull}
\end{figure}

To the extent that the gaussian estimate Eq.~\ref{eq:work1} is accurate,
the observed variance of binding energies observed for the
fully coupled system teaches us about available states of the filter
alone.  Fig.~\ref{fg:pushpull} illustrates this connection between the available
states for the uncoupled selectivity filter and the observed
fluctuations for the fully coupled system. 

We anticipate results that follow by noting that $\beta \sigma^2/
2$ takes values ranging from 12~kcal/mol to 6~kcal/mol, roughly, for the
cases of Na$^+$ to K$^+$.  In fact, $\bar\varepsilon$ will be
substantially lower for Na$^+$ than for K$^+$. The physical
interpretation from Eq.~\ref{eq:sep1} is that we would need  sample
sizes as big as
\begin{eqnarray}
\frac{1}{\int\limits_{-\infty}^{\bar{\varepsilon}}P_\mathrm{X}^{(0)}\left(
\varepsilon\right)\dif\varepsilon} 
 = \exp\left\lbrack{\beta\left(\mu^\mathrm{ex}_\mathrm{X} -
\left\langle\varepsilon\right\rangle\right)}\right\rbrack \nonumber \\
\approx\left\{ 
\begin{array}{c@{\quad}l}
5\times 10^8, &  \mathrm{(Na^+)}, \\
2\times 10^4 , & 
\mathrm{(K^+)}~
\end{array}
\right.
\label{spt-G}
\end{eqnarray}
to find a filter configuration that provides a binding energy
$\varepsilon\le\bar\varepsilon$ by sampling from uncoupled systems. If
we found one favorable configuration in such samples, the probabilities
would be all correct.  Configurations of the uncoupled filter that would be
favorable for Na$^+$ are thus less probable than those that would be
favorable for K$^+$.

As a check for the gaussian approximation Eq.~\ref{eq:work1}, we have also computed
$\Delta\mu^\mathrm{ex}$(filter) by transforming K$^+$ to Na$^+$ on the
basis of a coupling-parameter integration through 20 intermediate
states. This is, of course, a simple algorithmic approach to evaluation
of  Eq.~\ref{eq:pdt1}. Note that by introducing multiple gaussians
\cite{lrp:mulGjacs97}, the single gaussian approximation can be refined
to achieve quantitative agreement with coupling-parameter integration.
We do not pursue this point further here in favor of physical
clarity.

\section{Calculations}

The atomic coordinates for the protein structure were obtained from the
Protein Data Bank (PDB ID: 1BL8 \cite{rmk:science98}). Only residues 63
to 85 were retained in the simulation model (Fig~\ref{fg:model}).
Residues 74 to 79 comprise the selectivity filter. Following earlier
notation \cite{roux:nature04}, residues 76 (V) and 77 (G) comprise the
binding site denoted as S$_2$. All our calculations pertain to a single
ion located in the S$_2$ site.

The N-terminal group of each of the four protein chains was acetylated,
the C-terminal group was amidated, and all eight carboxylates were
ionized. Hydrogen atom positions were built and the structure was energy
minimized keeping only the non-crystallographically determined atomic
positions free. This initial structure provided the starting point for
further simulations. 

The heavy atoms outside the selectivity filter experience a harmonic
mean field with force constant 10~kcal/mole-{\AA}$^2$, and
similarly the heavy atoms of the selectivity filter felt harmonic
external forces corresponding to  $k$ =10, 5,
2.5~kcal/mole-{\AA}$^2$. We also considered a hybrid case in
which $k=0.0$ for the carbonyl oxygens of the S$_2$, but  $k=2.5$ for
all other atoms of the filter;  this is plotted as $k=0.0$ in
Fig.~\ref{fg:tab1fig}.

Molecular dynamics studies were carried out with the NAMD program
\cite{namd} using the CHARMM27 \cite{charmm}  forcefield. A temperature
of 298~K was maintained by velocity scaling. The  Lennard-Jones
parameters for the ions were from \cite{roux:jcpKNa}.  All the non-bonded
interactions were switched off from 17~{\AA} to 20~{\AA}.  A shorter
cutoff does not affect the results.

In using Eqs.~\ref{eq:pdt1} and~\ref{eq:work1}, we could consider the
ion-protein interaction in full.   But the binding energy $\varepsilon$
is particularly sensitive to local, near-neighbor interactions, and
far-field contributions are expected to be less discriminating between
Na$^+$ and K$^+$ in the filter.  Therefore, to obtain differences
$\Delta\mu^\mathrm{ex}\left(\mathrm{filter}\right)$ we compute only the
local interactions between the ions and ligands explicitly present. We
investigated several cases for including ion-filter interactions of
different types in evaluating the binding energy;  our quantitative
results changed by about 20\% in the extremes, and those different
approximations did not affect our conclusions below. For clarity then,
we present results for the case where interactions contributing to the
binding energy are limited to the carbonyl groups alone. Complementary
coupling-parameter integrations with the same scheme for considering
interactions were also performed.

For bulk hydration studies, the SPC/E \cite{spce} water model was used as
much of our preliminary work had been done with this water model. The
ion parameters had been developed for the TIP3P water model, but the
hydration results with SPC/E are not significantly different. The
experimental partial molar volume of the ion \cite{marcus} was used to
fix the simulation volume. For aqueous simulations, long range
electrostatics were treated using Ewald summation, with
non-electrostatic interactions cutoff  at 8.8~{\AA}. 

The coupling-parameter transformation of K$^+$ to Na$^+$ was carried out
in 20 steps, with 5~ps (10~ps for aqueous runs) for equilibration and
20~ps for statistical averaging. For calculations using the gaussian
model Eq.~\ref{eq:work1}, a simulation length of 0.5~ns was used and the
data stored every 20 steps (25 steps for aqueous runs) for analysis.
 
\section{Results and Discussion}

\subsection{Bulk hydration}
The bulk hydration free energies set the baseline for
filter selectivity  as is evident in Eq.~\ref{eq:selec}. The
coupling-parameter integration yields
$\Delta\mu^\mathrm{ex}\left(\mathrm{aq}\right)$ = $-$20.7, in good
agreement with \cite{roux:jcpKNa}.   The gaussian model gives
$-23.0$~kcal/mole. This result is reasonable for a single
gaussian description of aqueous hydration \cite{lrp:mulGjacs97}, and is consistent with the
coupling-parameter integration result.

From Eq.~\ref{eq:selec}, the relative selectivity is
$\Delta\mu^\mathrm{ex} =
\Delta\mu^\mathrm{ex}\left(\mathrm{filter}\right) + 20.7$~kcal/mole. It
is estimated that $\Delta\mu^\mathrm{ex}$ is about 6~kcal/mole
\cite{roux:nature04}, which is about three times smaller than the
difference in bulk hydration free energies. Thus the hydration
thermodynamic properties in bulk aqueous solution alone play a
dominating role in the filter selectivity.

\subsection{Filter results}

\begin{figure}
\begin{center}
\includegraphics[width=3in]{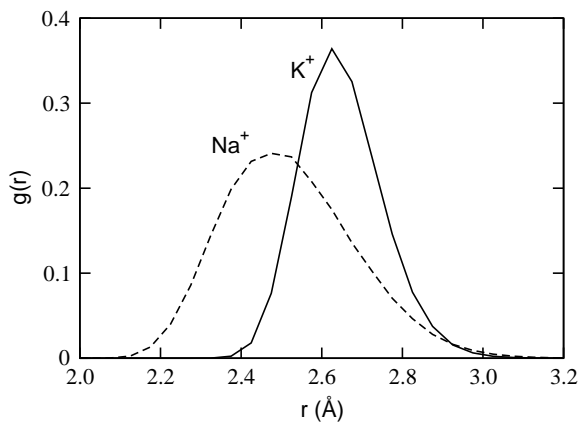}
\end{center}
\caption{Normalized distribution of carbonyl oxygens radially from the ion,
for the $k=0$ (hybrid) case described in  the text for the wild type (bold lines)
and the (GG) mutant (lighter lines). Notice that the longest
ion-oxygen distance is about the same in each case, but the shortest ion-oxygen
distances, and the most probable ones, are shorter in the case of Na$^+$.  
This suggests  that Na$^+$ isn't merely delocalized, but 
constricts the filter relative to the K$^+$ case.} \label{fg:gr}
\end{figure}

Fig.~\ref{fg:gr} shows the ion--carbonyl-oxygen pair distribution
function. Notice that carbonyl oxygen atoms  approach the Na$^+$ more
closely, consistent with the smaller size of Na$^+$ and the more
favorable coulombic interactions obtained then. The carbonyl oxygens are
also more {\em delocalized\/} in the case of Na$^+$ than that of K$^+$.
These observations imply that the selectivity filter has conformational
flexibility to readily accommodate Na$^+$, as opposed to the static
structural picture discussed in standard texts, for example Fig 13.25 in
\cite{berg:02} and Fig 11.24 in \cite{alberts:02}, in which the ion
moves about within a fixed filter structure. Instead these results
suggest that the Na$^+$ would achieve favorable binding energies by
shifting the ensemble of configurations for the filter away from that 
for the case of K$^+$.

\begin{figure}
\begin{center}
\includegraphics[width=3in]{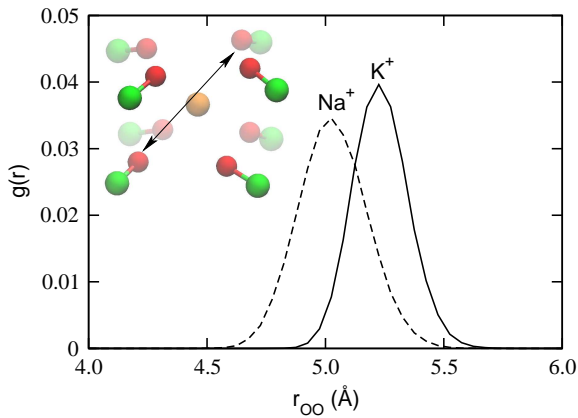}
\end{center}
\caption{Normalized radial distribution of `body-diagonal' OO pairs,
indicated by the arrow, for the selectivity filter in the cases of
Na$^+$ and K$^+$, indicating relative constriction of the filter by
Na$^+$. Bold lines: wild type. Lighter lines: (GG) mutant.  
Inset: atoms of the selectivity filter, showing an example of
the OO body-diagonal.  Carbonyl oxygen atoms are in red, carbonyl
carbons are green, and the ion is orange.  Notice that the width
of the distribution for the Na$^+$ case is slightly the larger, and 
that relative widths, $width/ r_\mathrm{max}$, of these distributions are
roughly 10\%.}\label{fg:carbonyl}
\end{figure}

This relative constriction can be directly observed as
Fig.~\ref{fg:carbonyl} shows.  This narrowing is expected to be
sensitive to the repulsive interactions between carbonyl-oxygen atom
pairs which have been implicated in selectivity \cite{roux:nature04}.
But selective manipulation of carbonyl OO repulsions is artificial and,
thus, experimental investigation of the necessity of this point does not
seem likely. 

\begin{figure}
\begin{center}
\includegraphics[width=3in]{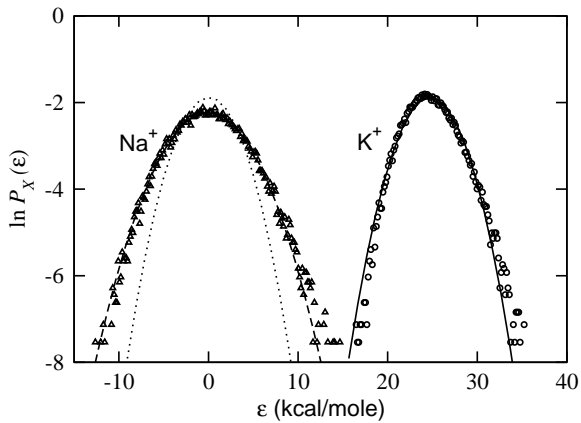}
\end{center}
\caption{Normalized distribution of ion carbonyl binding energies for the $k=0$ case for the
wild-type. The
circles (K$^+$) and triangles (Na$^+$) are the raw data. The binding
energy abscissa is measured relative the mean binding energy,
$\left\langle \varepsilon \right\rangle$, for the Na$^+$ case.   Thus,
the mean binding energy for K$^+$ is about 22-25~kcal/mol higher than
for Na$^+$. Gaussian fits to the data are smooth curves, and the
additional dotted line on the left superposes the fit for the K$^+$ case on
the Na$^+$ data. This shows that distribution is distinctly broader for
the Na$^+$ case. Similar results are obtained for the $k=10, 5.0, 2.5$
cases.
}\label{fg:psi}
\end{figure}

Fig.~\ref{fg:psi} shows the distribution of ion-carbonyl binding
energies for the $k=0$ case. It is clear that the mean interaction
energy of Na$^+$ with the filter is much lower than for K$^+$. In addition, the
distribution of this interaction energy for Na$^+$ is broader than that for K$^+$, consistent
with relative widths of the distributions of  Fig.~\ref{fg:gr}.

\begin{figure}
\begin{center}
\includegraphics[width=3in]{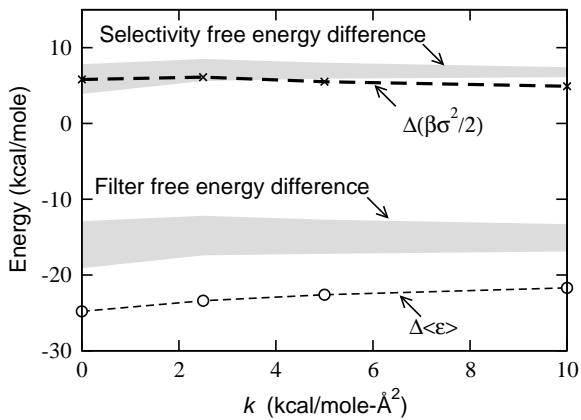}
\end{center}
\caption{Free energy results for several values of the force constant
$k$ used in the harmonic mean field.   The lower gray band is $\Delta\mu^\mathrm{ex}\left(\mathrm{filter}\right)$
of Eq.~\eqref{eq:selec}.  The  upper gray band 
is $\Delta\mu^\mathrm{ex}$, the left-side of Eq.~\eqref{eq:selec}.  The lower boundary of each grey
band is the result of the gaussian model Eq.~\ref{eq:work1}, and the
upper boundary was obtained on the basis of coupling-parameter
integration. Note that the mean interaction energy differences (circles)
are negative, signifying the better interaction of Na$^+$ with the
channel. }
\label{fg:tab1fig}
\end{figure}

Fig.~\ref{fg:tab1fig} collects the results for the various cases
considered. These results are not importantly sensitive to $k$. 
Consistent with experiments, $\Delta\mu^\mathrm{ex} > 0$: the filter
selects K$^+$ over Na$^+$. The magnitude of $\Delta\mu^\mathrm{ex}$ is
also roughly consistent with the estimate of about $5-6$~kcal/mole
\cite{roux:nature04}.

Nevertheless, Na$^+$ achieves the energetically more-favorable binding 
to the filter. $\Delta\left\langle\varepsilon\right\rangle$, the
difference in mean binding energy of Na$^+$ to the filter relative to
that for K$^+$, is between between $-22$~kcal/mole and $-25$~kcal/mole,
consistent with the energy difference noted in \cite{roux:nature04}.
These values are about the same as the difference in aqueous hydration
free energy of $-23$~kcal/mole. Thus, filter discrimination against
Na$^+$ on the basis of binding energies alone is implausible.

However, the difference in fluctuation contributions,
$\Delta(\beta\sigma^2/2) $, must also be considered, and the magnitude
of this difference is comparable to the net selectivity, as
Fig.~\ref{fg:tab1fig} shows. $\Delta(\beta\sigma^2/2) $ is roughly
$5-6$~kcal/mole, indicating that filter configurations conducive to
binding Na$^+$ are comparatively rare, consistent with Eq.~\ref{spt-G}. 
It is this difference in fluctuation contributions for Na$^+$ relative
to K$^+$ that shifts the balance in favor of K$^+$.

\section{Conclusions}

Na$^+$ binds strongly to the selectivity filter with a mean binding
energy substantially lower than for K$^+$.  The difference is comparable
to the difference between the hydration free energies for these two ions
in bulk aqueous solution.  Since the ion-sorting ability of the KcsA
K$^+$ filter must be considered from the baseline of this substantial
difference in bulk hydration free energies,  we conclude that the
average filter binding energies alone do not provide significant
discrimination of Na$^+$ from K$^+$.

Strong binding of the smaller Na$^+$ also constricts the selectivity
filter. From the point of view of the filter, discrimination against
Na$^+$ results from the fact that the observed Na$^+$-narrowed
conformations are rare occurrences in the ensemble of conformations
favorable to K$^+$. This effect is described in the thermodynamics as an
observed fluctuation contribution that is destabilizing for Na$^+$
relative to K$^+$.   The key result here is the association in the case
of Na$^+$ of strong binding, constriction of the filter, and enhanced
fluctuations.

With respect to the \emph{snug-fit} view of channel selectivity, it is
clear that a favorable binding energy is important.  In the case of the
smaller ion (Na$^+$), this is achieved at the expense of stronger
energetic and positional (Fig.~\ref{fg:carbonyl}) fluctuations.  For the
ion considered to have the better geometrical fit (K$^+$), this is
achieved at a lower cost in energetic and positional fluctuations.  In
this sense, the size difference between Na$^+$ and K$^+$, and the
molecular complementarity of the selectivity filter with K$^+$ do play a
role in channel selectivity.  From the \emph{fluctuations} point of the
view, the idea of a pre-configured empty channel is
hypothetical. Conformational flexibility induced by strong binding of
the smaller Na$^+$ in the selectivity filter is central to the mechanism
of Na$^+$ discrimination.
 
This work supports the view that selectivity can be addressed by
analysis of local interactions involving a single-ion binding site
\cite{roux:nature04}. The assessment of rates, though, likely requires
an account of interactions between ions occupying different channel
binding sites  \cite{ZhouYF:TheoiK}, but fluctuations of the sort
identified here probably play a role too.

The identification of channel-reorganization (induced fit) fluctuations is suggestive
of solvent reorganization in chemical dynamics. The fact that a simple
distribution, gaussian with slight positive skewness as in
Fig.~\ref{fg:psi}, works satisfactorily will have important conceptual
and practical consequences for modeling aqueous electrolyte solutions
more broadly.

\section*{Acknowledgements}
This work was supported by the US Department of Energy, contract
W-7405-ENG-36, under the LDRD program at Los Alamos. LA-UR-05-4622.
Financial support from the National Science Foundation (CTS-0304062) and
the Department of Energy (DE-FG02-04ER25626) is gratefully
acknowledged.


\clearpage

\section*{Figures}


\end{document}